\documentclass[twoside,draft,reqno,12pt]{amsart}
\usepackage[english]{babel}
\usepackage{amsmath}
\usepackage{amssymb}
\usepackage{enumerate}

\usepackage{amsfonts}
\usepackage{graphicx}

\usepackage[ citecolor=black, linkcolor=black,
colorlinks, hypertexnames=false, plainpages=false]{hyperref}

\usepackage{color}

\newtheorem{lemma}     {Lemma}[section]
\newtheorem{thm}   [lemma]{Theorem}
\newtheorem{teorema1}   [lemma]{Theorem}

\newtheorem{coro}       [lemma]{Corollary}
\newtheorem{cong1}      [lemma]{Conjecture}
\newtheorem{remark1}    [lemma]{Remark}
\newtheorem{defin}      [lemma]{Definition}

\numberwithin{equation}{section}

\newcommand{\R}{\mathbb R}
\newcommand{\Z}{\mathbb Z}

\newcommand{\dis}{\displaystyle}

\newcommand{\mmmintone}[1]{{\dis{\int\kern -.38cm
-}}_{\kern-.21cm\substack{#1}}\;\;}
\newcommand{\mmmintwo}[2]{{\dis{\int\kern -.43cm
-}}_{\kern-.21cm\substack{#1}}^{\substack{#2}}\;\;}
\newcommand{\submint}{{\scriptstyle{\int\kern -.66em -}}}
\newcommand{\submintone}[1]{{\scriptstyle{\int\kern -.66em
-}}_{\scriptscriptstyle{\kern-.21em\substack{#1}}}}
\newcommand{\fracmint}{{\textstyle{\int\kern -.88em -}}}
\newcommand{\fracmintone}[1]{{\textstyle{\int\kern -.88em
-}}_{\scriptscriptstyle{\kern-.21em\substack{#1}}}\;}

\newcommand{\ga}{\gamma}
\newcommand{\Ga}{\Gamma}

\newcommand{\La}{\Lambda}

\newcommand{\E}{\mathbb E}

\newcommand{\nada}[1]{}

\def\be{\begin{equation}}
\def\ee{\end{equation}}

\def\supp{{\rm supp}\,}

\def\La{\Lambda}

\def\GG{\mathcal{G}}
\def\CC{\mathcal{C}}
\def\BB{\mathcal{B}}
\def\TT{\mathcal{T}}

\def\Ra{{\rm R}}

\begin{document}
\today

\vskip.5cm
\title[Canonical ensemble]
{Cluster expansion in the canonical ensemble}

\author{Elena Pulvirenti}
\address{Elena Pulvirenti,
Dipartimento di Matematica, Universit\`a di Roma 3 \newline
\indent Rome, 00146, Italy}
\email{pulviren@mat.uniroma3.it}

\author{Dimitrios Tsagkarogiannis}
\address{Dimitrios Tsagkarogiannis,
Dipartimento di Matematica, Universit\`a di Roma Tor Vergata \newline
\indent Rome, 00133, Italy}
\email{tsagkaro@mat.uniroma2.it}

\begin{abstract}
We consider
 a system of particles confined in a box $\La\subset\R^d$ 
interacting via a tempered and stable pair
potential.
We prove the validity of the cluster expansion
for the canonical partition function
 in the high temperature -
low density regime.
The convergence is uniform in the volume
and in the thermodynamic limit 
it reproduces
Mayer's virial expansion
providing an alternative and more direct derivation which avoids the deep
combinatorial issues present in the original proof.
\end{abstract}

\maketitle


\section{Introduction}\label{sec1}

The quantitative prediction of macroscopic properties of matter through
its microscopic structure has been a main challenge for statistical mechanics.
In this direction a very important theoretical as well as practical
contribution is the work of J. E. and M. G. Mayer \cite{MM} (see also Ursell \cite{U}) 
in the theory of non-ideal
gases where they derive a full series expansion correcting the equation of state
for the ideal gas ($p=kT\rho$, where $p$ is the pressure of the system at temperature 
$T$ with density $\rho$, $k$ being the Boltzmann constant) to all orders in $\rho$ obtaining the famous {\it virial expansion}.
Convergence of this series has been addressed later (see \cite{LP1} and \cite{R}).
The main idea in \cite{MM}, (see also \cite{M})
is to describe all possible interactions among the particles of the non-ideal gas
by representing them as
linear graphs, which has later led to a main tool in statistical mechanics, 
namely the {\it cluster expansion} method. 
The thermodynamic
pressure is computed as the infinite volume limit of the logarithm of the
grand canonical partition function, which is however a function of the {\it activity} of the system.
To get an expansion with respect to the thermodynamic {\it density}, one needs to further 
express the latter
in a power series of the activity, invert it and replace it in the equation for the pressure.
This gives the virial expansion after using (for the inversion) 
some interesting combinatorial properties
of enumeration of graphs which finally express the coefficients of the virial expansion
to be sums over only ``irreducible graphs".

This road which leads to an expansion of the free energy versus the
density is evidently not the
straight one! The direct and natural way is to take the
density $\rho$ instead
of the activity
as order parameter and correspondingly to work in
the canonical rather than in
the grand canonical ensemble, which however rests on the possibility to cluster
expand   the canonical partition function.  We came to this problem
from other directions, see the end of the introduction, and found, to
our surprise, that the problem is not only solvable but easy. It fits
in fact beautifully in the theory of cluster expansion for abstract
polymer models, as  developed in all details by many authors
after the pioneering work of \cite{KP}, \cite{GK}. 
The exponentiation procedure in this
theory produces a lot of diagrams which are not present in the Mayer
expansion and which therefore must vanish in the thermodynamic limit.
As we shall see in Section~\ref{sec5} the origin of such cancellations is
closely related to the basic property of the cluster expansion (from
which the expansion takes its name) namely that the only chains of
graphs (clusters) which survive in the  expansion are made of ``incompatible''
graphs.

The validity of the cluster expansion for the canonical ensemble
opens the way to attack several other problems (which was actually
our initial motivation) such as the
finite volume corrections to the free energy, the radius of convergence of the expansion
in powers of the density (rather than the activity) for both the general model and
the particularly interesting case of hard spheres, 
the construction of coarse-grained Hamiltonians 
via multi-canonical constraints as required by the Lebowitz-Penrose Theorem \cite{LP2}
for Kac interactions
with applications to the LMP model \cite{LMP} and its variants.
We hope to further address these issues in subsequent papers.

\bigskip

\section{The model and the result}\label{sec2}

We consider a configuration ${\bf q}\equiv\{q_1,\ldots,q_N\}$ of $N$ particles in a box 
$\La:=(-\frac{L}{2},\frac{L}{2}]^d
\subset\R^d$ (for some $L>0$) 
which interact with a stable and tempered pair potential
$V:\R^d\to\R$, i.e., there exists $B\geq0$ such that:
\be\label{3}
\sum_{1\leq i<j \leq N} V(q_i-q_j) \geq -BN,
\ee
for all $N$ and all $q_1,...,q_N$
and the integral
\be\label{4}
C(\beta):= \int_{\R^d} |e^{-\beta V(q)}-1| dq
\ee
is convergent for some $\beta>0$ (and hence for all $\beta>0$).
Since in this paper we are interested in the infinite volume limit of the free energy
density, we can assume 
periodic boundary conditions since it
is a general result (see e.g. \cite{R} and \cite{FL}) 
that the thermodynamic limit is independent of the
choice of the boundary conditions.
This particular choice in the present paper is not essential, periodic boundary conditions
are used in order to obtain translation invariance in some cases (see e.g. Lemmas~\ref{lem1} and~\ref{l5.0}).
Furthermore,
our result remains valid with other boundary conditions by slightly changing the proof
and we hope to address it in a subsequent work 
where we will consider the finite volume corrections 
as well.

We obtain the periodic boundary conditions
by covering $\R^d$ with boxes $\La$
and adding all interactions. Let
\be\label{0}
V^{per}(q_i,q_j):=
 \sum_{n\in\Z^d} V(q_i-q_j+nL),
\ee
where apart from assuming
stability and temperedness, we further
need to guarantee convergence of the above sum by imposing a
condition on the decay properties of $V$.
A potential $V$ is called {\it lower regular} if there exists a decreasing
function $\psi:\R_+\to\R_+$ such that
$V(x)\geq -\psi(|x|)$ for all $x\in\R^d$ and
$\int_0^{\infty}\psi(s)s^{d-1}ds<\infty$.
Then $V$ will be called {\it regular} if it is lower regular and there
exists some $r_{V}<\infty$ such that $V(x)\leq\psi(|x|)$
whenever $|x|\geq r_{V}$ and this is our extra assumption (see also the discussion in \cite{FL}).

The {\it canonical partition function} of the system with periodic boundary conditions
is given by
\be\label{1}
Z^{per}_{\beta,\La,N}
:=\frac{1}{N!}\int_{\La^N} dq_1\,\ldots dq_N \,e^{-\beta H^{per}_{\La}(\bf q)},
\ee
where $H^{per}_{\La}$ is the energy of the system with periodic boundary conditions
given by
\be\label{2}
H^{per}_{\La}({\bf q})
:=\sum_{1\leq i<j \leq N} V^{per}(q_i,q_j).
\ee

Given $\rho>0$ we define the {\it thermodynamic free energy} by
\be\label{4.1}
f_{\beta}(\rho):=\lim_{|\La|,N\to\infty,\,\frac{N}{|\La|}=\rho} f_{\beta,\La}(N)
,\,\,\,{\rm where}
\,\,\,
f_{\beta,\La}(N)
:=
-\frac{1}{\beta|\La|}\log Z^{per}_{\beta,\La,N}.
\ee
The main result in this paper is:
\begin{thm}\label{thm1}
If $\rho\,C(\beta)$ is small enough (see \eqref{c2.1})
then 
\be\label{p3.1}
\frac{1}{|\La|}\log Z^{per}_{\beta,\La,N}=
\log\frac{|\La|^N}{N!}+\frac{N}{|\La|}\sum_{n\geq 1} F_{N,\La}(n),
\ee
where the coefficients $F_{N,\La}(n)$, $n\geq 1$, satisfy 
\be\label{n1}
|F_{N,\La}(n)|\leq C e^{-cn},
\ee
for constants $C,c>0$ that are independent of $N$ and $\La$.
Moreover,
\be\label{n5}
F_{N,\La}(n)=\frac{1}{n+1}P_{N,|\La|}(n)B_{\beta,\La}(n),
\ee
where $P_{N,|\La|}(n)$ and $B_{\beta,\La}(n)$ are defined in \eqref{p9}
and which in the thermodynamic limit give
\be\label{n3}
\lim_{N,|\La|\to\infty,\,N/|\La|=\rho}P_{N,|\La|}(n)=\rho^{n}\quad\text{and}\quad
\lim_{\La\to\infty}B_{\beta,\La}(n)=\beta_n,
\ee
for all $n\geq 1$, where $\beta_n$ are the irreducible coefficients of Mayer:
\be\label{n4}
\beta_{n}:=
\frac{1}{n!}
\sum_{\substack{g\in\BB_{n+1}\\ \supp g\ni\{1\}}}\int_{(\R^d)^n}\prod_{\{i,j\}\in E(g)}(e^{-\beta V(q_i-q_j)}-1)
dq_2\ldots dq_{n+1},\quad q_1\equiv 0
\ee
and $\BB_{n+1}$ is the set of $2$-connected (or irreducible) 
graphs $g$ on $(n+1)$ vertices
(i.e., connected graphs which by removing one vertex and all related edges remain
connected)
and $E(g)$ is the set of edges of the graph $g$.
\end{thm}
Note the unfortunate coincidence of notation between the inverse temperature $\beta$
and the irreducible coefficients $\beta_n$, which however we keep in agreement with the
literature.
To prove Theorem~\ref{thm1}
we first establish in Section~\ref{sec3}  the set-up of the cluster expansion for the canonical
partition function in the context of the abstract polymer model following \cite{BZ}, 
\cite{NOZ} and
\cite{KP}. Note that we could also work with more general formulations such as in
\cite{PU}, \cite{FP} or \cite{SS}.
In Section~\ref{sec4} we prove the convergence condition and as a corollary of
the cluster expansion theorem we prove \eqref{n1}.
The discussion of the thermodynamic limit is left for Section~\ref{sec5.0} 
where using the splitting in \eqref{n5} 
we prove \eqref{n3} by investigating the cancellations that lead to the
``irreducible" coefficients $\beta_n$, given in \eqref{n4}.
This latter fact is the result of some special structure (that we call ``product structure")
which may occur already in the abstract formulation of the polymer model and we address it in full generality in Subsection~\ref{sec5}.
Last, we also give an appendix with the main ideas of the original {\it virial} expansion
and connect it to our approach. There is a number of interesting combinatorial
issues that arise, many of which have been extensively studied in graph
theory. With no intention of being exhaustive 
we just refer to \cite{L} which is inspired by the virial
expansion.

\bigskip

\section{Cluster expansion for the canonical partition function}\label{sec3}

We view the canonical partition function
$Z^{per}_{\beta,\La,N}$ as a perturbation around the ideal case, hence
normalizing the measure by multiplying and dividing by $|\La|$ in \eqref{1}
we write
\be\label{5.1}
Z^{per}_{\beta,\La,N}= Z_{\La,N}^{ideal} Z_{\beta,\La,N}^{int},
\ee
where 
\be\label{5.2}
Z_{\La,N}^{ideal}:=\frac{|\La|^N}{N!}
\quad \text{and}\quad
Z_{\beta,\La,N}^{int}:=
\int_{\La^N} \frac{dq_1}{|\La|}\,\ldots \frac{dq_N}{|\La|} \,e^{-\beta H^{per}_{\La}(\bf q)}.
\ee
For $Z_{\beta,\La,N}^{int}$ we use the idea of Mayer in \cite{MM} 
which consists of developing $e^{-\beta H^{per}_{\La}(\bf q)}$ in the following
way
\be
e^{-\beta H^{per}_{\La}(\bf q)}=\prod_{1\leq i<j\leq N}(1+f_{i,j})
=\sum_{g\in\GG_N}\prod_{\{i,j\}\in E(g)}
f_{i,j},
\ee
where by $\GG_N$ we denote the set of
all simple graphs (undirected with no self-loops and no multiple edges) 
on $N$ vertices (the labels of the
particles), where a {\it graph} is the pair $g\equiv (V(g),E(g))$
where $V(g)\subset\{1,\ldots,N\}$ is the set of {\it vertices} and 
$E(g):=\{\{i,j\},i,j\in V(g)\}$ is the set of (non-oriented) {\it edges},
and
\be
f_{i,j}:=e^{-\beta V^{per}(q_i-q_j)}-1.
\ee

Given a graph $g\in\GG_N$ we say that it is {\it connected} if any two vertices of $g$
are connected by following its edges.
We define the support of a graph (denoted by $\supp g$) 
to be the set of its vertices.
Two graphs $g,g'$ are {\it compatible} (denoted by $g\sim g'$) 
if $\supp g\cap\supp g'=\emptyset$ (i.e., they are not connected); 
otherwise we call them {\it incompatible} ($\nsim$).
We also denote by $|g|$ the {\it cardinality} of $\supp g$.
With these definitions we see that any graph $g\in\GG_N$ is equivalent to
the pairwise compatible (non-ordered) collection of its connected components,
i.e. $g\equiv\{g_1,\ldots,g_k\}_{\sim}$ for some $k$. 
Hence,
\be\label{6}
Z_{\beta,\La,N}^{int}:=\sum_{\{g_1,...,g_k\}_{\sim}} \prod_{i=1}^{k}\tilde\zeta_{\La}(g_i)
=\sum_{\{V_1,...,V_k\}_{\sim}} \prod_{i=1}^{k}\zeta_{\La}(V_i),
\ee
where
\be\label{7}
\zeta_{\La}(V):= \sum_{g \in \CC_{V}}\tilde\zeta_{\La}(g),\qquad
\tilde\zeta_{\La}(g):=
\int_{\La^{|g|}} \prod_{i\in \supp g} \frac{dq_i}{|\La|}  \prod_{\{i,j\}\in E(g)} f_{i,j}.
\ee
Here $\CC_V$ denotes the set of all connected graphs with support in
$V\subset\{1,\ldots,N\}$. We will also be using the notation $\CC_n$ for the set of
all connected graphs on $n$ vertices. 
Both expressions in \eqref{6} are in the form of the abstract polymer model which
we specify next.

\medskip
An {\it abstract polymer model} ($\Ga$, $\mathbb{G}_{\Ga}$, $\omega$) 
consists of (i) a set of polymers 
$\Ga:=\{\ga_1,..., \ga_{|\Ga|}\}$, (ii) a binary symmetric relation $\sim$ of compatibility 
between the polymers (i.e., on $\Ga \times \Ga$) which is recorded into the
compatibility graph $\mathbb{G}_{\Ga}$
(the graph with vertex set $\Ga$ and with an edge between two polymers $\ga_i,\ga_j$ if and only if they are an incompatible pair)
and (iii) a weight function $\omega : \Ga \to \mathbb{C}$.
Then, 
we have the following formal relation which will become
rigorous by Theorem~\ref{thm2} below (see \cite{KP}, \cite{BZ} and \cite{NOZ}):
\be\label{poly1}
Z_{\Ga,\omega}:=\sum_{\{\ga_1,...,\ga_n\}_{\sim}} \prod_{i=1}^{n} \omega (\ga_i)=\exp\left\{
\sum_{ I\in \mathcal I} c_{ I}\omega^{ I}
\right\},
\ee
where
\be\label{7.2}
c_I=\frac{1}{I!}\sum_{G\subset\GG_I}(-1)^{|E(G)|},
\ee
or equivalently (\cite{BZ},\cite{D})
\be\label{7.21}
c_I= \frac{1}{I!} \frac{\partial^{\sum_{\ga}I(\ga)} \log Z_{\Ga,\omega}}{\partial^{I(\ga_1)} \omega(\ga_1) \cdots \partial^{I(\ga_n)} \omega(\ga_n)} \Big|_{\omega(\ga)=0}.
\ee

The sum in \eqref{poly1} 
is over the set $\mathcal I$ of all multi-indeces $I:\Ga \to\{0,1,\ldots\}$
with $\supp I:=\{\ga \in \Ga :\,I(\ga)>0\}$,
 $\omega^I=\prod_{\ga}\omega(\ga)^{I(\ga)}$ and
 $\GG_I$ is the graph with $\sum_{\ga\in\supp I} I(\ga)$ vertices induced from
 $\GG_{\supp I} \subset \mathbb{G}_{\Ga}$ by replacing each vertex $\ga$ by the complete graph on
 $I(\ga)$ vertices.

  Furthermore, the sum in \eqref{7.2} is over all connected subgraphs $G$ of $\GG_I$ spanning the whole set of vertices of $\GG_I$
 and $I!=\prod_{\ga\in\supp I} I(\ga)!$.
Note that if $I$ is supported on a compatible collection of polymers $\ga$ then
 $c_{I}=0$.
  
We state the general theorem as in \cite{BZ}, \cite{NOZ} but in a slightly different form.
Let 
\be\label{7.3}
L=L(\delta)=\sup_{x\in (0,\delta)}\left\{\frac{-\log(1-x)}{x}\right\}
=\frac{-\log(1-\delta)}{\delta},
\ee
where $\delta$ will be small and therefore $L=1+O(\delta)$. 
The optimal bound for the convergence radius
is beyond the scope of the present paper, however, we hope to come back to this issue,
also in the particular case of hard spheres, in a subsequent work.
\begin{thm}\label{thm2}[Cluster Expansion]
Assume that there are two 
non-negative functions $a,c:\Ga\to\R$ such that for any $\ga\in\Ga$,
$|\omega(\ga)|e^{a(\ga)}\leq\delta$ holds, for some $\delta>0$ small.
Moreover, assume that for any polymer $\ga'$
\be\label{7.4}
\sum_{\ga\nsim\ga'}|\omega(\ga)|e^{a(\ga)+c(\ga)}\leq \frac{1}{L}a(\ga'),
\ee
where $L$ is given in \eqref{7.3}.
Then, for any polymer $\ga'\in\Ga$ the following bound holds
\be\label{7.5}
\sum_{I:\,I(\ga')\geq 1} |c_I\omega^I| e^{\sum_{\ga\in\supp I}I(\ga) c(\ga)}\leq L|\omega(\ga')| e^{a(\ga')+c(\ga')},
\ee 
where $c_I$ are given in \eqref{7.21}.
\end{thm}

{\it Proof.} Apply the proof in \cite{BZ} for the activities
$\omega(\ga)e^{c(\ga)}$.\qed
  
  \bigskip
In view of \eqref{6} we can represent the partition function $Z_{\beta,\La,N}^{int}$
both as a polymer model on graphs with
($\CC(N)$, $\mathbb{G}_{\CC}$, $\tilde\zeta_{\La}$)
where we explicit the dependence on $N$ by denoting by $\CC(N)$ the set of all
labeled connected graphs on up to $N$ vertices
and as a 
polymer model on vertices, with ($\mathcal V(N)$,  $\mathbb{G}_{\mathcal V}$, $\zeta_{\La}$)
where $\mathcal V(N):=\{V:\,V\subset\{1,\ldots,N\}\}$.

\bigskip

\section{Convergence of the cluster expansion, proof of \eqref{n1}}\label{sec4}

In this section we check the convergence condition of Theorem~\ref{thm2}.
We work in the case of vertices which corresponds to the abstract polymer model
 ($\mathcal V(N)$,  $\mathbb{G}_{\mathcal V}$, $\zeta_{\La}$)
where $\mathcal V(N):=\{V:\,V\subset\{1,\ldots,N\}\}$. 
Then, as a corollary of Theorem~\ref{thm2} we prove \eqref{n1}.

\begin{lemma}\label{lem1}
There exist two positive functions $a,c:\mathcal V(N) \rightarrow \mathbb{R}$
such that for any $V\in\mathcal V(N)$:
\be\label{conv}
|\zeta_{\La}(V)|e^{a(V)}\leq\delta
\ee
holds for some $\delta>0$ small.
Moreover, for any $V'\in\mathcal V(N)$
\be\label{c0}
\sum_{V: \text{ }V \not\sim V'} |\zeta_{\La}(V)| e^{a(V)+c(V)} \leq \frac{1}{L}a(V'),
\ee
where $L$ is given in \eqref{7.3}.
\end{lemma}

\emph{Proof}.
We choose $a(V)=a |V|$ 
and $c(V)=c |V|$, for some $a, c>0$ to be chosen later and let $\alpha:=a+c$.
To bound $|\zeta_{\La}(V)|$ 
we use a version of the tree-graph inequality (proved in this form
in \cite{PU}
and \cite{R}) which states that
for a stable and tempered potential, we have the following bound:
\be
\Big|\sum_{g \in \CC_{n}} \prod_{\{i,j\} \in E(g)} f_{i,j}\Big| \leq 
e^{2\beta B n}\sum_{T \in\TT_n} \prod_{\{i,j\}\in E(T)} |f_{i,j}|,
\ee
where $\TT_n \subset \CC_n$ is the set of trees with $n$ vertices. Then
\begin{eqnarray}
|\zeta_{\La}(V)| e^{a |V|} &\leq & e^{(2\beta B+a)|V|}\int_{\La^{|V|}} \prod_{i\in V} \frac{dq_i}{|\La|}\sum_{T \in\TT_{|V|}} \prod_{\{i,j\}\in E(T)} |f_{i,j}| \nonumber\\
&\leq &
e^{(2\beta B+a)n} \sum_{T \in\TT_n} \int_{\La^n} \frac{dq_1}{|\La|} \cdots \frac{dq_n}{|\La|} \prod_{\{i,j\}\in E(T)} |f_{i,j}|\label{c1}.
\end{eqnarray}
Given a rooted tree $T$ let us call $(i_1,j_1),(i_2,j_2),...,(i_{n-1},j_{n-1})$ its edges.
We have:
\begin{align} 
\int_{\La^n} \frac{dq_{1}}{|\La|} \cdots \frac{dq_{n}}{|\La|} \prod_{\{i,j\} \in E(T)} | f_{i,j} |  &=
\frac{1}{{|\La|}^n}\int_{\La^n} dq_{1}\cdots dq_{n} \prod_{k=1}^{n-1} | f_{i_k,j_k} |  \notag \\
&
\leq \frac{1}{{|\La|}^n}\int_{\La} dq_{i_1} \int_{\La} dy_2\cdots  \int_{\La} dy_{n} \prod_{k=2}^{n} | e^{ -\beta V^{per}(y_k)}-1| \notag \\
&
\leq \frac{|\La|}{{|\La|}^{n}}\left[ \int_{\La} dx | e^{ -\beta V^{per}(x)}-1|\right]^{n-1}
\notag=:\frac{|\La|}{|\La|^n}C_{\La}(\beta)^{n-1},
\end{align}
(note that the choice of $V^{per}$ makes $C_{\La}(\beta)$ independent of $x$)
where we considered $q_{i_1}$ as the root and we used the change of variables:
\be
y_k= q_{i_k}-q_{j_k}, \qquad \forall k=2,...,n. 
\ee
We choose $\rho\, C(\beta)$ such that: 
\be\label{c2.1}
\rho e^{2\beta B+\alpha+1}C(\beta)<1.
\ee
Then, since the number of all trees in $\TT_n$ is $n^{n-2}$,
from \eqref{c1} we obtain (recalling that $N/|\La|=\rho$):
\be\label{c2}
|\zeta_{\La}(V)| e^{a |V|} \leq
\frac{n^{n-2}}{|\La|^{n-1}}e^{(2\beta B+a)n}C_{\La}(\beta)^{n-1}\leq
\frac 12 
\rho\, C_{\La}(\beta)e^{2(2\beta B+a)},
\ee
by using the bound $2\leq n\leq N$ and the fact that $\rho e^{2\beta B+a}C_{\La}(\beta)<1$. 
The latter is true considering that inequality \eqref{c2.1} still holds 
with $C_{\La}(\beta)$ for $\La$ large enough, since 
$\lim_{\La\to\infty}C_{\La}(\beta)=C(\beta)$.
Then defining $\delta:=\frac 12 
\rho\, C(\beta)e^{2(2\beta B+a)}$,
\eqref{conv} is satisfied.
Considering $a=1$ we have
(for any fixed $i$):
\be\label{c3}
\sum_{V: \text{ }V \ni i} |\zeta_{\La}(V)| e^{\alpha |V|} \leq
\sum_{n\geq 2} \binom{N-1}{n-1}
\frac{n^{n-2}}{|\La|^{n-1}}e^{(2\beta B+\alpha)n}C_{\La}(\beta)^{n-1}\leq e\delta\frac{1}{1-\delta'}\leq \frac 1L ,
\ee
recalling that $L=1+O(\delta)$
and where $\delta':= \rho e^{2\beta B+\alpha+1}C(\beta)$.
Since $\{V \not \sim V' \}\subset \bigcup_{i\in V'} \{
V\ni i\}$ we get \eqref{c0} and conclude the proof of the lemma.\qed

\medskip
The way we chose to present the cluster expansion as well as its convergence
can by no means give the best radius of convergence.
Our goal was merely to obtain (giving up the seek for the best radius) the consequence
of the cluster expansion theorem, given in \eqref{7.5}, 
which we use in order to establish \eqref{p3.1}.
Nevertheless, our condition is comparable with the ones in the literature 
(see \cite{LP1}, equation (3.15), and in \cite{R}, Theorem 4.3.2) and we will
clarify these issues in a subsequent work.
Moreover, 
for the particular case of the hard spheres we can
obtain the improved radius as in \cite{FPS} but for the density $\rho$ 
rather than the activity,
since we are working directly with the canonical partition function.

\bigskip
{\it Proof of \eqref{n1}.}
After proving the convergence condition it follows directly from Theorem~\ref{thm2}
that for all $V'\in\mathcal V(N)$ and by choosing $c(V):=c |V|$ and $a(V):=|V|$
\be\label{cor}
\sum_{I:\,I(V')\geq 1} |c_I\zeta_{\La}^I| e^{c\| I\|}\leq L |\zeta_{\La}(V')| e^{\alpha|V'|},
\quad
\|I\|:=\sum_{V\in\supp I}I(V) |V|,
\ee
where we remind that
$\alpha=1+c$. 
Thus, exponentiating the partition function \eqref{6} using \eqref{poly1},
we obtain \eqref{p3.1}:
\be\label{p0}
\frac{1}{|\La|}\log Z^{int}_{\beta,\La,N} 
=\frac{1}{|\La|}\sum_{I} c_{I}\zeta_{\La}^{I} 
=\frac{1}{|\La|} \sum_{i\in\{1,...,N\}} \sum_{I : \supp I \ni \{i\}} \frac{1}{| I |}c_{I}\zeta_{\La}^{I} 
=
\frac{N}{|\La|}
\sum_{n\geq 1}
F_{N,\La}(n),
\ee
where
\be\label{p7}
F_{N,\La}(n):= \frac{1}{n+1}
\sum_{\substack{I:\,\supp I\ni \{1\} \\ |I|=n+1}}
c_{I}\zeta_{\La}^{I},\quad |I|:=\left|\bigcup_{V\in\supp I}V\right|.
\ee
Note that $|I|\leq\|I\|$.
The function $F_{N,\La}(n)$ is uniformly bounded for all $N,\La$
as well as absolutely summable over $n$, namely from \eqref{cor} with 
$V'\equiv\{1\}$ we get:
\be\label{cor0}
|F_{N,\La}(n)|
\leq 
\frac{e^{-cn}}{n+1}\sum_{\substack{I:\,\supp I\ni \{1\} \\ |I|=n+1}} |c_{I}\zeta_{\La}^{I}|e^{cn}
\leq
e^{-cn}Le^{\alpha},
\ee
which concludes the proof of \eqref{n1}.\qed

\medskip
Note that the sum in \eqref{p7} is infinite. As it will be apparent in the next section 
it is more convenient to
work with a truncated version of $F_{N,\La}(n)$. Thus, 
for $M\geq n+1$ we define
\be\label{cor2}
F^M_{N,\La}(n):=\frac{1}{n+1}
\sum_{\substack{I:\,\supp I\ni \{1\} \\ |I|=n+1,\,\|I\|\leq M}}
c_{I}\zeta_{\La}^{I},
\ee
which is uniformly bounded for all $N,\La$
and absolutely summable over $n$ using the same argument as in \eqref{cor0}.
Note also that now the sum is finite.
Furthermore, for all $n\geq 1$
\be\label{cor3}
\sup_{N,\La}| F^M_{N,\La}(n)-F_{N,\La}(n)|
\leq 
\sup_{N,\La} \frac{e^{-\frac c2(n+M)}}{n+1}
\sum_{\substack{I:\,\supp I\ni \{1\} \\ |I|=n+1,\, \|I\|>M}} 
|c_{I}\zeta_{\La}^{I}|e^{\frac c2(n+M)}
\leq
e^{-\frac c2(n+M)}Le^{\alpha} ,
\ee
since inside the sum we have: $n\leq M<\| I \|$.
Hence, we can next work with the truncated function $F^M_{N,\La}(n)$ 
and then pass simultaneously to both limits: the thermodynamic and $M\to\infty$.

\bigskip

\section{The thermodynamic limit, proof of \eqref{n3}}\label{sec5.0}

The sum in the definition of $F^M_{N,\La}(n)$ \eqref{cor2} does not depend on the 
labels of the extra $n$ particles (we have already chosen label $1$). Hence,
\be\label{p8}
F^M_{N,\La}(n)=\frac{1}{n+1}\binom{N-1}{n}
\sum_{\substack{I:\,\|I\|\leq M \\ \supp I\equiv\{1,\ldots,n+1\}}}
c_{I}\zeta_{\La}^{I}=\frac{1}{n+1}P_{N,|\La|}(n)B^M_{\beta,\La}(n),
\ee
where
\be\label{p9}
P_{N,|\La|}(n):=\frac{(N-1)\ldots(N-n)}{|\La|^n}\quad\text{and}\quad
B^M_{\beta,\La}(n):=\frac{|\La|^n}{n!}\sum_{\substack{I:\,\| I \| \leq M \\ \supp I\equiv\{1,\ldots,n+1\}}}
c_{I}\zeta_{\La}^{I}.
\ee
While obviously $P_{N,|\La|}(n)\to\rho^n$, for $B^M_{\beta,\La}(n)$ 
we proceed as follows:
we have restricted ourselves to a system of only $n+1$ particles, therefore we can
view the sum in $B^M_{\beta,\La}(n)$ as the cluster sum in
($\mathcal V(n+1),\mathbb G_{\mathcal V},\zeta_{\La}$) with the extra condition 
$\|I\|\leq M$.
While for the convergence it was more convenient to work with the space ($\mathcal V(N),\mathbb G_{\mathcal V},\zeta_{\La}$),
the cancellations that we discuss in this section occur already at the more elementary
structure of labeled connected graphs 
(i.e., on ($\CC(n+1),\mathbb G_{\CC},\tilde\zeta_{\La}$) whose multi-indeces we denote
by $\tilde I:\CC(n+1) \to\{0,1,\ldots\}$) 
so since
$\sum_{I:\,\|I\|\leq M} c_I\zeta_{\La}^I=\sum_{\tilde I:\,\|\tilde I\| \leq M} c_{\tilde I}\tilde \zeta_{\La}^{\tilde I}$
(coming from the same partition function)
we rewrite $B^M_{\beta,\La}(n)$ as follows:
\be\label{p1}
B^M_{\beta,\La}(n)=\frac{|\La|^n}{n!}
\sum_{\tilde I:\,\|\tilde I\|\leq M} c_{\tilde I}\tilde \zeta_{\La}^{\tilde I}
=\frac{|\La|^n}{n!}\sum_{g\in\CC_{n+1}}
\sum_{\substack{\tilde I:\,\|\tilde I\|\leq M \\ \cup_{g'\in\supp\tilde I}g'=g}} c_{\tilde I}\tilde\zeta_{\La}^{\tilde I}.
\ee
Note that looking at the support of some cluster $\tilde I$ we see a connected graph
(with possible superposition of some graphs $g$ if $\tilde I(g)\geq 2$, or with some common 
parts of a graph
if they are incompatible in more than one points). Thus, in \eqref{p1} 
we first put in evidence the resulting graph $g\in\CC_{n+1}$
and then sum over all $\tilde I$ that correspond to it.
The main idea of what follows is to investigate the order in $|\La|$ of the products 
$\tilde\zeta_{\La}^{\tilde I}$. We start with a definition:

\begin{defin}\label{def1}
Given a connected undirected graph a vertex is said to be an {\it articulation point} if removing it and all edges incident to it the graph results in a non-connected graph.
\end{defin}

One of the main ingredients of our proof is the fact that if $\{g':\,g'\in\supp\tilde I\}$ 
are connected
at only articulation points of the fixed $g\in\CC_{n+1}$ then the activities $\tilde\zeta_{\La}$ factorize.
To emphasize this fact we state it as a lemma:
\begin{lemma}\label{l5.0}
Given $g\in\CC_{n+1}$ and an incompatible collection $G$ such that
$\cup_{g'\in G}g'=g$, if for all incompatible pairs $g',g''$ we have that
$\supp g'\cap\supp g''=\{i\}$ is a singleton and moreover, the vertex $i$
is an {\it articulation point} of $g$
then
\be\label{p1.10}
\tilde\zeta_{\La}(g)=\prod_{g'\in G}\tilde\zeta_{\La}(g'),
\ee
for all finite $\La$.
\end{lemma}

{\it Proof.} 
By the definition of an articulation point there is a $g'\in G$ which has only one articulation point. 
We integrate the coordinates in $g'$ keeping fixed the one that corresponds to the articulation point. In principle the result should depend on the position 
of the articulation point, but this is ruled out by the special choice of boundary conditions 
(as we stressed in Section~\ref{sec2})
which makes such integrals translation invariant.
Hence, the result of the integration gives $\tilde\zeta_{\La}(g')$.
Then, again by the definition
of an articulation point,
this can continue until integrating all $g'\in G$.\qed

\bigskip
Coming back to \eqref{p1}
the power of $|\La|$ is
$n-\sum_{g'\in\supp\tilde I}(|g'|-1)\tilde I(g')$.
Moreover, since it is always true that 
$n+1\leq \sum_{g'\in\supp\tilde I}(|g'|-1)+1$
(by the fact that all $g'\in\supp\tilde I$ should be incompatible, i.e., have at least one
common label)
 it is 
 implied that non-negligible terms (in the limit $\La\to\infty$) should satisfy
\begin{eqnarray}
&&\tilde I(g')=1,\,\forall g'\in\supp \tilde I,\,\,\, \text{and}\label{p1.3}\\
&& n+1= \sum_{g'\in\supp\tilde I}(|g'|-1)+1\label{p1.4}.
\end{eqnarray}
(In the representation of hyper-graphs by bi-partite graphs, these two requirements 
correspond to graphs without loops.)
Again, to emphasize we state it as a lemma
\begin{lemma}\label{l5.01}
Given $\tilde I$, if there is $A\subset\supp\tilde I$ such that
$\tilde\zeta_{\La}(\cup_{g'\in A}g')\neq \prod_{g'\in A}\tilde\zeta_{\La}(g')$ then the cluster that
corresponds to $\tilde I$ is vanishing
at the thermodynamic limit.
\end{lemma}

{\it Proof.} 
Suppose that there is such a
$A\subset\tilde I$, i.e., 
\be\label{p1.21}
\left |\bigcup_{g'\in A}g'\right |\neq\sum_{g'\in A}(|g'|-1)+1.
\ee
Then, \eqref{p1.21} implies that
\be\label{p1.22}
n+1=\left|\bigcup_{g'\in\supp\tilde I}g'\right|\leq\left|\bigcup_{g'\in A}g'\right|
+\left|\bigcup_{g'\in \supp\tilde I\setminus A}g'\right|-1
<\sum_{g'\in A}(|g'|-1)+1+\sum_{g'\in\supp\tilde I\setminus A}(|g'|-1)+1-1,
\ee
hence $\eqref{p1.4}$ cannot be true and it has to vanish by taking the limit.\qed

\medskip

{\it Proof of \eqref{n3}.}
Motivated by these facts we proceed with \eqref{p1} by splitting the sum over
$\tilde I$ into the following
two complementary cases:
\begin{enumerate}
\item\label{it1}
(factorization property holds)
 $\tilde\zeta_{\La}(\cup_{g\in A}g)=\prod_{g\in A}\tilde\zeta_{\La}(g)$ for all $A\subset\supp\tilde I$,
\item there exists $A\subset\supp\tilde I$ such that $\tilde\zeta_{\La}(\cup_{g\in A}g)\neq\prod_{g\in A}\tilde\zeta_{\La}(g)$.
\end{enumerate}
Lemma~\ref{l5.01} states that in the limit $|\La|\to\infty$ the terms
of Case $(2)$ vanish (and they are finitely many due to the truncation).
For the case $(1)$, if $g\in\CC_{n+1}\setminus\BB_{n+1}$ then all terms 
in $B^M_{\beta,\La}(n)$ 
are exactly
canceling each other.
This is a property of the combinatorial coefficients
multiplying products of activities of polymers which have the factorization property,
just as in the cluster expansion where compatible collections (carrying the factorization property) do not contribute in the sum.
In the next subsection we investigate this property that
we call ``product structure" (see Definition~\ref{def1}) and since it can 
be a general property of the abstract polymer model we prove it in its general form 
in Lemma~\ref{l5.1} and then we explicit it in our case in
Corollary~\ref{l5.2}.

Thus, the only remaining terms are those of Case 1 for $g\in\BB_{n+1}$ which due to
condition \eqref{p1.3} give non-vanishing contribution only for $\tilde I(g)=1$ (and those with higher multiplicity are only finitely many, due to the truncation). Hence, 
for all $n\geq 1$ we can pass to the thermodynamic limit 
(note that \eqref{cor3} permits to exchange sum and limits)
and obtain:
\be\label{f4}
\lim_{\La\to\infty}B_{\beta,\La}(n)=
\lim_{\La\to\infty}\lim_{M\to\infty}B^M_{\beta,\La}(n)=\lim_{\La\to\infty}\frac{|\La|^n}{n!}\sum_{g\in\BB_{n+1}}\tilde\zeta_{\La}(g)=\beta_n,
\ee
where $\beta_n$ is defined in \eqref{n4},
concluding \eqref{n3} and
the proof of Theorem~\ref{thm1} (pending the proof of Corollary~\ref{l5.2}).\qed

\bigskip
\subsection{The ``product structure"}\label{sec5}

Coming back to the general formulation ($\Ga,\mathbb G_{\Ga},\omega$),
for any $\Ga'\subset\Ga$ we define  
the set of all incompatible sequences
that can be constructed out of elements of $\Ga'$ by:
\be\label{s5.0}
\Ga'_{\nsim}:=\{A:\, A\subset\Ga',\,\text{incompatible}\}.
\ee
Recall that every single element $\ga\in\Ga'$ is considered incompatible,
hence the singleton 
$\{\ga\}$ is an element in $\Ga'_{\nsim}$). 

\begin{defin}\label{def2}
Given $(\Ga, \mathbb{G}_{\Ga},\omega)$,  we say that $\Ga^b\subset\Ga$ 
has a ``product structure" if
\begin{itemize}
\item there exists a one-to-one function $\phi:\Ga^b_{\nsim}\to \Ga$,
with $\phi(\{\ga\})=\ga$, if $\ga\in\Ga^b$,
\item for any $A\in \Ga^b_{\nsim}$,
we have the factorization
\be
\omega(\phi(A))=\prod_{\ga'\in A}\omega(\ga').
\ee
\end{itemize} 
\end{defin}
We also define the {\it range} of $\phi$ by
\be\label{range}
\Ra(\phi):=\{\phi(A),\forall A\in\Ga^b_{\nsim}\}.
\ee
We are interested in all multi-indeces $I$ such that 
every $\ga\in\supp I$ 
is the
image of some $A\in\Ga^b_{\nsim}$,
i.e.,
$\supp I\subset\Ra(\phi)$ 
and this is the content
of the following lemma.
\begin{lemma}\label{l5.1}
If $\Ga^b\subset\Ga$ has a product structure
then in the expansion \eqref{poly1} we have:
\be\label{poly4}
 \sum_{I:\, \supp I\subset\Ra(\phi)} c_{ I}\omega^{ I}=\sum_{I:\,\supp I\equiv\{\ga'\},\,\ga'\in\Ga^b}c_I\omega^I.
\ee
\end{lemma}

\medskip
{\it Proof.}
Given $\Ga^b\subset\Ga$ with product structure, using \eqref{range}, we define
\begin{eqnarray}\label{s5.1}
Z^*(\Ga^b)& := & Z_{\Ga,\{\omega(\ga)\equiv 0,\,\forall \ga\notin \Ra(\phi)\}}
\equiv\sum_{\substack{\{\ga_1,...,\ga_k\}_{\sim} \\ \ga_i \in \Ra(\phi)}}
\prod_{i=1}^k\omega(\ga_i)\nonumber\\
& = &
\sum_{\substack{\{A_1,...,A_k\}_{\sim} \\ \phi(A_i)=\ga_i,\,\forall i}}
\prod_{i=1}^k\omega(\phi(A_i))
=\prod_{\ga'\in\Ga^b}(1+\omega(\ga')).
\end{eqnarray}
The first equality of \eqref{s5.1} 
is due to the fact that $\phi$ is one-to-one, i.e., 
for any $\ga_i\in \Ra(\phi)$
there is a unique $A_i\in\Ga^b_{\nsim}$ with $\phi(A_i)=\ga_i$.
Then using
the factorization property, i.e., 
$\omega(\phi(A_i))=\prod_{\ga'\in A_i}\omega(\ga')$,
the second equality is due to the fact that
$\prod_{\ga'\in\Ga^b}(1+\omega(\ga'))
=\sum_{A\subset\Ga^b}\prod_{\ga'\in A}\omega(\ga')$,
where the latter sum is over subsets $A$ (compatible or incompatible) 
of the set of vertices in $\Ga^b$, 
hence it can be uniquely decomposed into $k$ compatible components 
$A\equiv\{A_1,\ldots,A_k\}_{\sim}$ 
with $A_i\in\Ga^b_{\nsim}$ for all $i$.

Then if we exponentiate the last expression of \eqref{s5.1} we obtain the right hand side
of \eqref{poly4}, while if we exponentiate 
$Z_{\Ga,\{\omega(\ga)\equiv 0,\,\forall \ga\notin \Ra(\phi)\}}$ 
(by first exponentiating using \eqref{7.21} and then evaluating)
we obtain the left hand side 
of \eqref{poly4}.\qed

\bigskip

Now we explicit the result of Lemma~\ref{l5.1} for the case of \eqref{p1}, which
corresponds to the particular
polymer model ($\CC(n+1), \mathbb G_{\CC},\tilde\zeta_{\La})$.
Any $g\in\CC_{n+1}\setminus\BB_{n+1}$ can be uniquely decomposed into $2$-connected components,
say $g=\cup_{i=1}^k b_i$ where $b_i\in\BB(n+1),\, \forall i=1,\ldots,k$
(recall the difference of notation between $\BB_{n+1}$ and $\BB(n+1)$ with the latter
being the set of $2$-connected graphs on {\it up to} $n+1$ vertices).
Then, according to Definition~\ref{def2}, any such element $\{b_1,\ldots,b_k\}$
has product structure with
$\phi(A):=\cup_{b\in A}b$ for all $A\subset\{b_1,\ldots,b_k\}$ incompatible
and it also satisfies (see Lemma~\ref{l5.0}) the factorization property by construction.
Thus,
\begin{coro}\label{l5.2}
Given $g\in\CC_{n+1}\setminus\BB_{n+1}$, then
\be\label{cor1}
\sum^*_{\substack{\tilde I:\,\|\tilde I\|\leq M\\ \cup_{g'\in\supp\tilde I}g'=g}} 
c_{\tilde I}\tilde\zeta_{\La}^{\tilde I}=0,
\ee
for all truncations $M$ and
where $*$ is to remind that the sum is over $\tilde I$ who are supported on polymers $g$
which have the factorization property, as in Case~\ref{it1} in the proof of \eqref{n3}.
\end{coro}

{\it Proof.}
Suppose that $\{b_1,\ldots, b_k\}$ is the unique decomposition into 
$2$-connected
components of $g$ and hence has product structure. 
Then applying Lemma~\ref{l5.1} for 
$\Ga^b:=\{b_1,\ldots, b_k\}$ 
we obtain that all coefficients of 
$\tilde\zeta_{\La}(b_1)^{n_1}\ldots\tilde\zeta_{\La}(b_k)^{n_k}$
(with $n_1|b_1|+\ldots+n_k|b_k|\leq M$)
are zero except those with $k=1$, which however correspond to the case 
$g=b_1\in\BB_{n+1}$ and are not considered since
we assumed $g\in\CC_{n+1}\setminus\BB_{n+1}$.\qed

\bigskip

\section{Appendix: Mayer's virial expansion}

The approach introduced by Mayer, see \cite{MM}, is to work in the grand canonical ensemble 
which is a measure on both the number of particles $N$ and the configuration ${\bf q}$
\be\label{4.31}
G_{\beta,z,\La}(N;d{\bf q}):=\frac{1}{\Xi_{\beta,\La}(z)}z^N  \,e^{-\beta H_{\La}(\bf q)}
dq_1\,\ldots dq_N,
\ee
where $\Xi_{\beta,\La}(z)$ is the grand canonical partition function given by
\be\label{4.5}
\Xi_{\beta,\La}(z):=\sum_{N\geq 0}z^N Z_{\beta, \La, N}
\ee
and $z$ is the {\it activity} of the system.
The thermodynamic pressure is defined as the infinite volume limit of the logarithm of the grand canonical partition function:
\be\label{4.4}
p_{\beta}(z):=\lim_{|\La|\to\infty} p_{\beta,\La}(z),\quad\text{where}\,\,\,
\beta p_{\beta,\La}(z)=\frac{1}{|\La|}\log\Xi_{\beta,\La}(z).
\ee

The idea in \cite{MM} consists of developing $e^{-\beta H_{\La}(\bf q)}$ in the following
way
\be\label{4.01}
e^{-\beta H_{\La}(\bf q)}=\prod_{1\leq i<j\leq N}(1+f_{i,j})
=\sum_{g\in\GG_N}\prod_{\{i,j\}\in E(g)}
f_{i,j},
\ee
where by $\GG_N$ we denote all simple graphs on $N$ vertices, 
$E(g)$ is the set of edges
of a graph $g\in\GG_N$ and
\be\label{5}
f_{i,j}:=e^{-\beta V(q_i-q_j)}-1.
\ee
Then the grand canonical partition function becomes
\be\label{4.51}
\Xi_{\beta,\La}(z)=\sum_{N\geq 0}\frac{z^N}{N!}
\sum_{g\in\GG_N} w_{\La}(g),\quad\text{where}\,\,\,\,
w_{\La}(g):=\int_{\La^{|g|}}\prod_{\{i,j\}\in E(g)}f_{i,j}
\prod_{i=1}^{|g|} dq_i,
\ee
where by $|g|$ we denote the cardinality of the graph $g$ and we define it to be the number of vertices.
Using the fact that the weight $w_{\La}(g)$ 
is multiplicative on disconnected components
a general theorem on enumeration of graphs gives (see e.g. \cite{UF} where it is stated as ``The first Mayer Theorem")
\be\label{4.52}
\sum_{N\geq 0}\frac{z^N}{N!}
\sum_{g\in\GG_N} w_{\La}(g)=
\exp\left\{
\sum_{n\geq 1}\frac{z^n}{n!}
\sum_{g\in\CC_n} w_{\La}(g)
\right\},
\ee
where $\CC_n$ is the set of {\it connected} graphs on $n$ vertices.
This is the predecessor of the Cluster Expansion Theorem~\ref{thm2}!
Then defining
\be\label{4.61}
b_{n}(\La):=\frac{1}{|\La| n!}\sum_{g\in\CC_n} w_{\La}(g)
\ee
(which is normalized in the volume and hence it has a limit $b_n:=\lim_{|\La|\to\infty} b_n(\La)$),
equation
\eqref{4.4} gives
\be\label{4.6}
p_{\beta,\La}(z)=\frac{1}{\beta |\La|}\sum_{n\geq 1} |\La| b_n(\La) z^n\to
\frac{1}{\beta}\sum_{n\geq 1} b_n z^n\equiv p_{\beta}(z).
\ee

In the thermodynamic limit the canonical free energy is the Legendre-Fenchel transform
of the pressure, namely
\be\label{4.62}
\beta f_{\beta}(\rho)=\sup_{z}\{
\rho\log z-\beta p_{\beta}(z)
\}=\rho\log z(\rho)-\beta p_{\beta}(z(\rho)),
\ee
where $z(\rho)$ is given by the inversion of the relation $\rho=z p_{\beta}'(z)$.
Note that this is also equivalent to first defining the finite volume density by
\be\label{4.7}
\rho_{\La}(z):=\E_{G_{\beta,z,\La}}[N]=z  p_{\beta,\La}'(z)
\ee
and then pass to the limit $|\La|\to\infty$.
In \cite{UF} this inversion
is referred as ``The second Mayer Theorem" and it is again a result on enumerating
connected and $2$-connected graphs, where the latter means all graphs which
cannot be reduced to connected graphs by removing a point and all
related edges.
Under again the assumption that $w_{\La}(g)$ is multiplicative (see in our case Lemma~\ref{l5.0}),
we have that
\be\label{4.9}
(\rho= z p_{\beta}'(z)=)z \frac{\partial}{\partial z}\left(
\sum_{n\geq 1}\frac{z^n}{n!} 
\sum_{g\in\CC_n} w(g)\right)
=z \exp\left\{
\frac{\partial}{\partial \rho}\left(
\sum_{m\geq 2}\frac{\rho^m}{m!}\sum_{g\in\BB_m} w(g)\right)
\Big|_{\rho:\, z=z(\rho)}
\right\}
\ee
where $\BB_m$ is the set of $2$-connected graphs on $m$ vertices.
Note that this is the combinatorial counterpart of our discussion in Section~\ref{sec5.0}.
From \eqref{4.9} we have that
\be\label{4.8}
\rho=z p_{\beta}'(z)\Leftrightarrow
z(\rho)=\rho e^{-\sum_{m\geq 2} \beta_{m-1}\rho^{m-1}
},
\ee
where
\be\label{4.3}
\beta_m:=\lim_{|\La|\to\infty}
\frac{1}{|\La| m!}\sum_{g\in\BB_{m+1}}
w_{\La}(g).
\ee
Plugging into $p_{\beta}(z(\rho))$ we obtain the famous {\it virial} expansion:
\be\label{4.10}
\beta p_{\beta}(\rho)=\rho-\sum_{m\geq 1}\frac{m}{m+1}\beta_m\rho^{m+1}.
\ee
Overall, \eqref{4.62}, gives
\be\label{4.2}
f_{\beta}(\rho)=\frac{1}{\beta}\left\{
\rho(\log\rho-1)-\sum_{m\geq 1}\frac{1}{m+1}\beta_m\rho^{m+1}
\right\}.
\ee

 \vskip1cm

 {\bf Acknowledgments.}

It is a great pleasure to thank Errico Presutti for suggesting us the problem discussed
in this paper and for his continuous advising.
We further acknowledge very fruitful discussions with Joel Lebowitz, Roman
Koteck{\'y}, Marzio Cassandro, Enzo Olivieri, 
Suren Poghosyan, Daniel Ueltschi and Benedetto Scoppola.
D.T. also acknowledges very kind hospitality of the 
Center for Theoretical Study at Prague 
and the Mathematics Institute of the University of Warwick.
The research of D.T. has been partially supported by a Marie Curie Intra European Fellowship within the 7th European Community Framework Program.

\end{document}